# Modeling the Spread of COVID-19 in University Communities

**Jeffrey W. Herrmann[1,2,*], Hongjie Liu[3], and Donald K. Milton[4]**

[1]  Institute for Systems Research, University of Maryland, College Park, MD 20742
[2]  Department of Mechanical Engineering, The Catholic University of America, Washington, DC 20064
[3]  Department of Epidemiology and Biostatistics, University of Maryland, College Park, MD 20742
[4]  Maryland Institute for Applied Environmental Health, University of Maryland, College Park, MD 20742

*  **Correspondence:** herrmannj@cua.edu

**Abstract:** Mathematical and simulation models are often used to predict the spread of a disease and estimate the impact of public health interventions, and many such models have been developed and used during the COVID-19 pandemic.    This paper describes a study that systematically compared models for a university community, which has a much smaller but more connected population than a state or nation.    We developed a stochastic agent-based model, a deterministic compartment model, and a model based on ordinary differential equations.    All three models represented the disease progression with the same susceptible-exposed-infectious-recovered (SEIR) model.    We created a baseline scenario for a population of 14,000 students and faculty and eleven other scenarios for combinations of interventions such as regular testing, contact tracing, quarantine, isolation, moving courses online, mask wearing, improving ventilation, and vaccination.    Where possible, our study used parameter values from other epidemiological studies and incorporated data about COVID-19 testing in College Park, Maryland, but the study was designed to compare modeling approaches to each other using a synthetic population, so comparisons with data about actual cases were not relevant.    For each scenario we used the models to estimate the number of persons who become infected over a period of 119 days (17 weeks in a semester).    We evaluated the models by comparing their predictions and evaluating their parsimony and computational effort.    The agent-based model (ABM) and the deterministic compartment model (DCM) had similar results with cyclic flow of persons to and from quarantine, but the model based on ordinary differential equations failed to capture these dynamics.    The ABM's computation time was much greater than the other two models' computation time.    The DCM captured some of the dynamics that were present in the ABM's predictions and, like those from the ABM, clearly showed the importance of testing and moving classes on-line.

## 1. Introduction

Public health officials and university officials have responded in multiple ways to the spread of the novel coronavirus (SARS-CoV-2). Their response has included various interventions such as closing residence halls and suspending research, instructional, and support activities on campus, mandating social distancing, requiring vaccinations, mandating or encouraging wearing masks, modifying ventilation systems in critical buildings, allowing vulnerable faculty and staff to work at home, mandatory testing for SARS-CoV-2, isolation, and contact tracing.

These interventions are meant to protect students, faculty, and staff by reducing the spread of COVID-19 in the university community. But they disrupt operations, utilize human resources, and cost money. Thus, university decision-makers benefit from having estimates of the effectiveness of different interventions in their community. To that end, disease spread models can provide valuable information. Models are abstractions of the real world that create an artificial "laboratory" for considering alternatives, such as which interventions to implement [1].

Consequently, during the pandemic, researchers have created numerous models of university communities to estimate the spread of the disease in those communities and the potential impact of different interventions. Ghaffarzadegan [2] used differential equations to model a hypothetical university with 25,000 students and 3,000 faculty and staff over a period of 90 days. This simulation study considered different policies such as social distancing, proactive testing, quarantine, mask use, and remote work for faculty and staff.

Bahl *et al.* [3] used an agent-based model to represent a small residential college with 2,000 students and 380 faculty in three schools. Their simulation study considered school policies (weekly screening tests and reducing the number of students living on campus) and student behavior (testing, wearing masks, and reduced social events). Gressman and Peck [4] used an agent-based model a university with 20,000 students and 2,500 instructors over a period of 100 days. Their simulation study considered testing, contact tracing, quarantine, moving large classes online, and wearing masks.

Benneyan *et al.* [5] used two models (one with differential equations and an agent-based model) to represent a university over a 16-week semester. Their simulation study considered the uncertainty in the parameter values by running the models repeatedly with different samples of the parameter values. They considered three settings that ranged from 42,000 to 110,000 persons (students and nearby residents). Goyal *et al.* [6] used an agent-based model to simulate the University of California San Diego; their model had over 38,000 students and 8,000 faculty and staff. Xi and Chan [7] used an agent-based model to simulate a university with 2,592 students who are living in four residence halls. This study considered the impact of self-quarantine for returning students, wearing masks, minimizing movement, and more frequent cleaning. Cator *et al.* [8] used an agent-based model to simulate a 15-week semester at a synthetic college community with 3,000 persons, and they used this model to estimate the impact of different vaccination, testing, and isolation interventions on an outbreak of COVID-19.

Paltiel *et al.* [9] used a deterministic compartment model of 5,000 students over 80 days. This study considered regular testing at different frequencies. Rennert *et al.* [10] used a deterministic compartment model of a university with 25,000 students over 90 days. Their study considered the impact of testing returning students in scenarios with different values for R0, which represent the effectiveness of non-pharmaceutical interventions. Rennert *et al.* [11] used a deterministic compartment model of a university with 18,548 students. Their study considered different surveillance testing strategies. Rennert *et al.* [12] used a deterministic compartment model with six compartments to model an on-campus population of 7,500 and evaluated different reopening and pre-arrival testing policies.

Cashore *et al.* [13] developed a stochastic compartment model of a campus with 20,000 students over 16 weeks. Their study considered testing returning students, quarantine, contact tracing, and regular testing.

These studies generally modeled the progression of the disease using the susceptible-exposed-infectious-recovered (SEIR) framework that has been used many times for modeling infectious diseases generally [14, 15] and COVID-19 in particular.

More generally, Weeden and Cornwell [16] used transcript data to determine the connections between over 13,500 college students via the approximately 3,800 courses in which they were enrolled. They found that students were enrolled in an average of 5.3 classes each, and these courses created chains of potential infection.

In this area, many different types of models have been created, including agent-based models (ABM), deterministic compartment models (DCM), and models with ordinary differential equations (ODE). Numerical solutions to ODE models, which have been used widely for modeling disease spread, can be found quickly, but such models are often based on simplifying assumptions. A DCM may allow more details to be included, but it also groups persons into compartments, which reduces the computational effort. An ABM allows one to model each person with as much detail as desired; many ABMs are stochastic, so that multiple replications are needed to estimate the expected disease spread.

Ajelli *et al.* [17] modeled a hypothetical pandemic and used two models to predict the spread of the disease in Italy. They used a detailed agent-based model and a stochastic metapopulation model, which incorporates compartment models for each urban area and uses data from a global mobility dataset for predicting the spread around the country. The models considered nearly 57 million Italians over a period of 350 days. Their metapopulation model generated larger estimates of the number of infected persons for the entire country and for each urban area, but the predicted peaks were just a few days apart. Their study, like the one described here, focused on the differences generated by different modeling approaches; comparing the simulation results with data about a real epidemic was not relevant.

Formulating a model requires developing an abstraction that makes assumptions about the real world in order to focus on the relationships that are essential for making decisions (such as which interventions to implement). Although the general tradeoffs between increasing model detail (and accuracy) and computational effort are well-known, we are unaware of any studies that have studied this tradeoff precisely for the case of COVID-19 in a university community. Although many researchers have developed models for an entire state, a nation, and the whole planet, a specific, relatively limited scenario is valuable because it complements such large-scale models and is able to model particulars that are of concern to local decision-makers.

This paper describes a study that we conducted to generate more specific insights into the modeling tradeoffs that exist in this domain. In particular, we created models of the university community to predict disease spread over one semester, and we studied the differences between the models' predictions, their inputs, and their computational requirements. These inter-model differences (not comparisons with the actual disease spread) were key to our study.

We built an ABM, a DCM, and an ODE model (ODEM). We created twelve scenarios that represent a baseline scenario and combinations of five interventions, including surveillance testing, reducing in-person courses, wearing masks, improving classrooms, and vaccination. The ABM required more computational effort than the DCM and ODEM. We used, as much as possible, the same parameter values and assumptions in all three models. The models' predictions showed that some interventions were more effective than others. The predictions from the ABM and DCM showed cyclic behavior that the predictions from the ODEM did not.

The key contribution of this study is the side-by-side comparison of three common modeling approaches for predicting disease spread using the case study of a synthetic but realistic university community. As the COVID-19 pandemic continues and in future pandemics, if any, these results will help university decision-makers and analysts make informed choices about which models to use to estimate the impact of interventions in their communities.

The remainder of this paper is organized as follows: Section 2 presents the approach that we followed in this study, including the different interventions that were considered, the methods for evaluating the models, the different models that we created, and how we modified them to model the interventions. Section 3 presents the results. Section 4 discusses the results. Section 5 concludes the paper. Appendices A and B present the equations for the DCM and ODEM.

## 2. Materials and methods

Our study included the following steps:
1. Define the baseline scenario;
2. Construct the models of the baseline scenario (ABM, DCM, and ODEM);
3. Identify the intervention scenarios;
4. Modify each model for each intervention scenario;
5. Run models for all scenarios and compare predictions.

Details about the baseline scenario (Step 1) and the intervention scenarios (Step 3) are included in the following subsections. Sections 2.3, 2.4, and 2.5 and Appendices A and B describe the models that we constructed and how we modified them for the interventions (Steps 2 and 4). Sections 3 and 4 present and discuss the results (Step 5).

Although none of the scenarios that were considered represent exactly what happened on the University of Maryland in the Fall 2020 semester, we used our university's actions to inform our design of the baseline and intervention scenarios so that they would be reasonable approximations of the options that were considered. Moreover, because detailed data about campus conditions that semester were not available to our research team, this study used a synthetic population that approximates but does not match precisely the university community. Because we were considering possible scenarios and studying the differences between the models' predictions, their inputs, and their computational requirements, this study did not attempt to model the actual conditions on campus or compare the models' predictions with the actual disease spread. (If adequate data were available, such a retrospective study might be possible.) When selecting values for key parameters about the pandemic, we found different modeling studies used different values. Here, we relied on studies that took similar approaches and considered similar questions. Most importantly, we used the same values in all of our models.

*2.1. Baseline Scenario*

The baseline scenario was the key scenario in our study; we modeled interventions by modifying the models of the baseline scenario, which was meant to represent the situation with the least restrictions and interventions.

In this scenario, the population has 14,000 persons. There are 12,000 students who live on campus and 2,000 faculty who visit campus to teach their courses. The 12,000 students live in two-person rooms, so each one has one roommate. (We did not consider students who live off campus.) The time horizon is 119 days (August 20 to December 16, 2020). Initially, three persons are infectious.

During the Fall 2020 semester, the University of Maryland required testing every two weeks (14 days) for anyone who was on campus. According to the university's COVID-19 dashboard [18], the greatest number of tests administered during any 14-day period was 14,006, and three positive test results were recorded on August 19, 2020. This suggested that approximately 14,000 people were on-campus during this time and that three persons were infectious at the beginning of the semester. (Students and faculty can also become exposed from interactions with the off-campus community; see Section 2.3 for details.)

There are 2,000 courses being offered in person, and each student takes five courses. Half of the courses meet on Monday and Wednesday; the other half meet on Tuesday and Thursday. The average class size is 30 students. Each member of the faculty teaches exactly one course. If the course is being held on campus, all students and faculty who are not in isolation or quarantine attend class. There are no other educational, recreational, or social events on campus. Anyone who becomes infectious undergoes testing and enters isolation if they receive a positive test result. Everyone complies with the rules about isolation and quarantine.

The baseline scenario represents the situation with no interventions and thus should describe the natural (unmitigated) dynamics of the spread of the disease in the university setting.

In all scenarios and models, we used a SEIR model to represent the progression of SARS-CoV-2. A person starts as susceptible (S). When a susceptible person contacts someone who is infectious (I), the susceptible person might become infected. Any susceptible persons who become infected are considered "exposed" (E); this state is also known as "latent infected." (In this study, the term "exposed" is consistent with the typical use of the SEIR model; it does not mean someone who was merely in the proximity of someone infected.) An exposed person is infected but not infectious; the person may be considered "effectively exposed" and "pre-infectious." The mean time in the exposed state is three days [9]. An exposed person will eventually become infectious; the mean time in the infectious state is fourteen days [9]; after being infectious, the person is recovered (R). In our models, which consider only one semester, a recovered person cannot become infected again.

When a susceptible person contacts an infectious person, the infectious person might transmit the virus to the susceptible person. The probability of transmission depends upon numerous factors; these will be covered in the description of the ABM (Section 2.3).

There are many details that are not included in this baseline scenario, and the models make many simplifications about the progression of COVID-19, the university community, and other factors. We included details that exemplify the types of factors that detailed models can represent, but some relevant data was not available, and this study's comparison of the models did not require including all possible details. Because this modeling effort began in 2020, we used parameter values for the original strain that was circulating in 2020. (It would be important to use updated parameter values for the most widespread strain when using the models.)

## 2.2. Interventions

Among the many possible interventions that public health officials have used in their attempts to slow the spread of COVID-19, we considered the following because they were implemented or considered at the University of Maryland:

• Testing: all persons are tested regularly (every 14 days). This intervention includes case isolation (anyone who tests positive must remain home and avoid contacts with others), contact tracing, and quarantine (10 days).

• Masks: anyone infectious who wears a mask reduces the probability of transmitting the virus to someone who is susceptible; anyone susceptible who wears a mask reduces the probability of being exposed during a contact with someone who is infectious. In scenarios with mask wearing, 25% of persons wear tight-fitting masks, 25% wear loose-fitting masks, and 50% wore no masks. (This might underestimate the number who wore masks, which would reduce the impact of mask wearing.) The impact of wearing different types of masks (see Section 2.2) was based on studies of mask wearing [19, 20].

• Moving most classes on-line: students who do not go to class will not have contacts with faculty and classmates (except for roommates). In the Fall 2020 semester, the University of Maryland provost stated that "full in-person instruction will only be at about 20%" [21].

• Environmental improvements: if the ventilation system in a particular building is improved, the probability of transmitting the virus for contacts in that building is reduced.

• Vaccination: Anyone susceptible who has been vaccinated reduces the probability of becoming infected during a contact with someone who is infectious. Anyone vaccinated who does become infected has a different distribution of the time to recovery. Four of the scenarios considered set the vaccination rate at 80%, which is close to the rate at the University of Maryland in July 2021 [22].

We then developed eleven intervention scenarios, which yielded a total of twelve scenarios. As shown in Table 1, these include scenarios with just one intervention and scenarios with different combinations of interventions. Half of the scenarios included testing. The descriptions in Section 2, Appendix A, and Appendix B provide details about how the models were modified to represent the interventions.

**Table 1.** Scenarios used in modeling experiments.　See text for details of each intervention.

| Scenario | Testing? | On-campus courses | Mask wearing | Cleaner classrooms | Vaccination rate |
|---|---|---|---|---|---|
| 1. Baseline | No | 100% | None | None | 0% |
| 2. Baseline with testing | Yes | 100% | None | None | 0% |
| 3. Move classes online | No | 20% | None | None | 0% |
| 4. Move classes online with testing | Yes | 20% | None | None | 0% |
| 5. Require mask wearing | No | 100% | 50% | None | 0% |
| 6. Require mask wearing with testing | Yes | 100% | 50% | None | 0% |
| 7. Cleaner classrooms | No | 100% | None | 100% | 0% |
| 8. Cleaner classrooms with testing | Yes | 100% | None | 100% | 0% |
| 9. Require vaccinations | No | 100% | None | None | 80% |
| 10. Require vaccinations with testing | Yes | 100% | None | None | 80% |
| 11. Require mask wearing and vaccinations, cleaner classrooms | No | 100% | 50% | 100% | 80% |
| 12. Require mask wearing and vaccinations, cleaner classrooms with testing | Yes | 100% | 50% | 100% | 80% |

All of the models were implemented and run using MATLAB R2020b on a Dell XPS 8930 with an Intel Core i7-8700 CPU running at 3.20 GHz with 32.0 GB RAM.　The details of the DCM and the ODEM are provided in Appendices A and B.

*2.3. Stochastic Agent-Based Model*

The stochastic agent-based model (ABM) simulates the spread of COVID-19 through the population by modeling each person's behavior each day and determining whether that person becomes infected that day.　The population has 14,000 persons: 12,000 students and 2,000 faculty.　The students are divided into 6,000 roommate pairs.　The simulation begins with three infectious persons; these are drawn at random from the students and faculty.

Each person who is not quarantined or isolated has contacts during the day.　The model assumes that each person has four "general" contacts; these four are selected at random from the entire population.　If the person is a student, then one of these four is the person's roommate.　The other contacts represent persons met while performing ancillary activities such as visiting the library, going to office hours, shopping, dining, and religious services.　If a student has a class that meets that day, his contacts also include everyone who attends class.

In this model, each student's five courses are selected at random from the list of 2,000 courses (the model does not group students into majors that make certain courses more likely than others).　Thus, the average class size is 30 students, but the actual number will vary by replication.　(In one replication, for instance, the class sizes ranged from 9 to 48 students.)

On any day that a course meets on campus, the class instructor and all students who are not in isolation or quarantine attend class, so they all contact each other that day.　(If the instructor is in isolation or quarantine, the class meets without him.)

For any two persons, the model saves the date of the last contact, which is needed for contact tracing.

A person who is susceptible may become infected (exposed) during the day due to a contact with someone on campus who is infectious or from a contact with someone off-campus.

Although the values of the following variables may change during the simulation, to simplify this notation, we do not include any index for the day here.

Consider person $i$ who is susceptible on day $t$. Let $C$ be the set of infectious persons whom person $i$ contacts that day. Let $v_i = 1$ if person $i$ is vaccinated and 0 otherwise. Let $c_v$ be the coefficient that reduces the probability of transmission due to vaccination. Let $m_j$ be the coefficient that reduces the probability of transmission due to the mask that person $j$ wears; $m_j = 1$ if person $j$ wears no mask; $m_j = 0.5$ if person $j$ wears a loose-fitting mask; $m_j = 0.1$ if person $j$ wears a tight-fitting mask. Let $p$ be the base probability of transmission from a contact with one infectious person. Let $y_{ij} = 1$ if persons $i$ and $j$ met that day in a "clean" classroom (and 0 otherwise), and let $c_c$ be the coefficient that reduces the probability of transmission due to a "clean" classroom.

Let $\bar{n}$ be the average number of positive cases in the community over the last five days, and let $p^o$ be the baseline value for the probability of transmission from the off-campus community.

For each infectious contact $j$ in $C$, the model calculates the probability of transmission, which equals $c_c^{y_{ij}} c_v^{v_i} m_i m_j p$, and then randomly selects from the corresponding Bernoulli distribution. If the sample equals 1, the person is infected and is considered "exposed." If none of his contacts infects person $i$, then the model calculates the probability of transmission from the community, which equals $\bar{n} p^o$ and then randomly selects from the corresponding Bernoulli distribution. If the sample equals 1, the person is infected and is considered exposed.

The ABM randomly selects the duration until the person is infectious and the duration until the person recovers. Each duration is selected from an exponential distribution; the means of these distributions were discussed in Section 2.1.

A person who is exposed remains in that state until the infectious period begins. A person who is infectious remains in that state until he recovers. A person who has recovered cannot become infected again and never again enters quarantine or isolation. (Although breakthrough infections can occur, our models disregard this possibility within the short time horizon of one semester.)

Because the ABM is stochastic, we ran ten replications for each scenario.

**Testing.** In scenarios with surveillance testing and contact tracing, each day, the probability that a person who is eligible for testing will be tested is 1/14. Anyone in quarantine is tested every day. Anyone who has a positive test result is placed into isolation the same day. On the day after the positive test result, anyone who had contact with him within the previous five days is placed into quarantine. For example, if Joe has a positive test result on day 11, Joe is put into isolation that day, and, on day 12, anyone whom he contacted on or after day 6 goes into quarantine. Isolation and quarantine begin at the beginning of the day, so those persons have no contacts that day and any days that they are in isolation or quarantine. Those who are in isolation exit when they have recovered (and have a negative test result); the quarantine time was set to ten days.

**Fewer on-campus courses.** In scenarios with fewer courses that meet on campus, a random set of courses was chosen at the beginning of the simulation run, and those courses never met on campus, so the students and faculty in those courses did not contact each other.

**Masks.** In scenarios in which persons wore masks, a random set of persons was chosen at the beginning of the simulation run, and the probability of transmission was reduced for any contacts with those persons.

**Environmental improvements.** In scenarios with ventilation improvements, a random set of classrooms was chosen at the beginning of the simulation run, and the probability of transmission was reduced for any contacts between persons in those classrooms.

**Vaccination.** In scenarios in which persons were vaccinated, a random set of persons was chosen at the beginning of the simulation run, and the probability of transmission was reduced for those persons.

*2.4. Deterministic Compartment Model*

The deterministic compartment model (DCM) and the ODEM are similar; neither one keeps track of the state of a person; instead, both models are deterministic and have compartments that represent groups of similar persons. A key difference between these two models is that the DCM model can represent factors that change each day, whereas the parameters in the ODEM that we implemented are invariant. The DCM uses difference equations, but the ODEM uses differential equations.

The DCM included twenty compartments, ten for students and ten for faculty:

- Susceptible (S)
- Exposed (E)
- Infectious (I)
- Recovered (R)
- Susceptible in isolation (SI)
- Exposed in isolation (EI)
- Infectious in isolation (II)
- Susceptible in quarantine (SQ)
- Exposed in quarantine (EQ)
- Infectious in quarantine (IQ)

Figure 1 shows the flows between compartments. The SQ compartment contains those who had a contact (with someone who had a positive test result) but were not actually infected (not "effectively exposed"); therefore, these persons are not in the E compartment.

After setting the initial values for every compartment, the model iterated over each day in the time horizon. For each day it calculated the expected number of persons who moved from one compartment to the other that day. Let $S$ be the set of all twenty compartments. Let $n_a(t)$ be the number of persons in compartment $a$ at the end of day $t$. Let $f_{ab}(t)$ be the number of persons who move from compartment $a$ to compartment $b$ during day $t$. $f_{aa}(t) = 0$ for all $a$ in $S$.

$$n_a(t) = n_a(t-1) + \sum_{b \in S} f_{ba}(t) - \sum_{b \in S} f_{ab}(t)$$

The equations for $f_{ab}(t)$ are included in Appendix A.

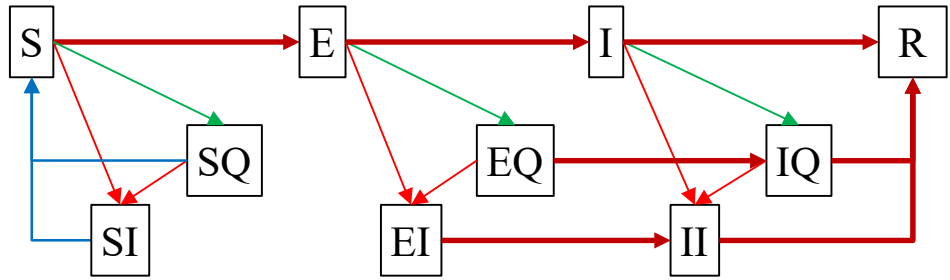

**Figure 1.** The deterministic compartment model has two groups, with ten compartments for students and ten compartments for faculty. This figure shows the flow pattern between the compartments for each group; students and faculty have the same flow pattern. Thick red arrows represent flows due to disease progression (SEIR); green arrows represent those who enter quarantine due to contacts; thin red arrows represent those who are isolated due to positive test results; blue arrows represent susceptible who leave isolation or quarantine.

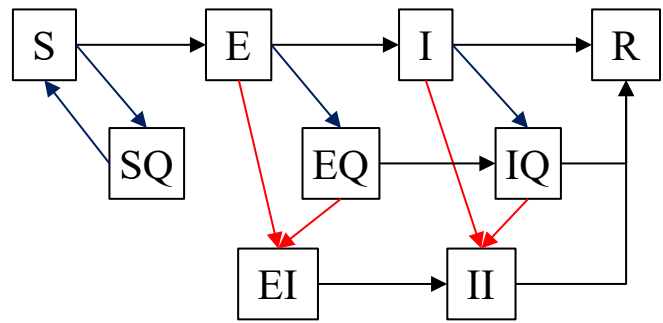

**Figure 2.** The ODEM has nine compartments, and the flow between compartments is indicated by the arrows in this figure. The red arrows indicate flows that depend upon positive test results.

## 2.5. Ordinary Differential Equations Model

The ordinary differential equations model (ODEM) was defined using a set of differential equations. This model had nine compartments:

- Susceptible (S)
- Exposed (E)
- Infectious (I)
- Recovered (R)
- Exposed in isolation (EI)
- Infectious in isolation (II)
- Susceptible in quarantine (SQ)
- Exposed in quarantine (SQ)
- Infectious in quarantine (SQ)

Figure 2 shows the flows between compartments. This model did not separate faculty and students; this was done to make the model even simpler than the DCM, which was useful for comparing the different models. This model did not account for false positives (thus, no susceptible persons were isolated). Appendix B lists its equations and describes how their parameter values were determined.

## 2.6. Student Cohorts

We repeated our approach with a different set of assumptions. We assumed that on-campus, outside-class contacts are more limited: students contact only students, and faculty contact only faculty. Moreover, students were randomly placed into cohorts; all students in the same cohort take the same classes; they take no classes with students from other cohorts. We modified the ABM and DCM to correspond to these changes; no change to the ODEM was needed.

## 3. Results

We ran each model (ABM, DCM, and ODEM) in every scenario. Table 2 lists the total number infected by scenario and model. We created figures that show the trajectory (over 119 days) of the number of persons who were exposed, in isolation, and in quarantine (Figures 3 to 5) and a figure of the number of new cases each week by scenario and model (Figure 6).

## 3.1. Predictions

The predictions for the total number of infected are listed in Table 2. The predictions of the number exposed, the number in isolation, and number in quarantine for all twelve scenarios are shown in Figures 3 to 5. The weekly number infected are shown in Figure 6.

As shown in Figure 6, the predicted weekly number infected remains elevated throughout the time horizon in Scenario 1, when no interventions are used. The pattern in other scenarios is different.

For scenarios with testing, contact tracing, and quarantine, the predictions from the ABM and the DCM show large, periodic changes in the number of persons in isolation and in quarantine. These occur because, in these models, the contact tracing identifies a large number of contacts (most of whom are susceptible), these contacts enter quarantine, and ten days later they exit quarantine and re-enter the susceptible category. When many susceptible persons are in quarantine, there are fewer susceptible persons who can be contacts, so the number of susceptible persons who enter quarantine decreases. Eventually, they exit quarantine, and the cycle begins again.

The models' predictions show that the intervention with the most impact is testing (scenarios 2, 4, 6, 8, 10, and 12). Testing includes contact tracing and quarantine, and quarantine reduces the number of contacts and disease spread. As shown in Figures 4 and 5, among these six scenarios, the number of persons in isolation and quarantine is much different for scenario 4, which has both testing and reducing the number of courses meeting on campus. When only 20% of the courses are meeting on campus, the number of contacts decreases, and this changes the trajectory of the infection. The other five scenarios with testing do not include reducing the number of courses meeting on campus, and their results are very similar because the impact of the other interventions is much smaller.

The predictions by the ABM and DCM for the six scenarios without testing (scenarios 1, 3, 5, 7, 9, and 11) are more similar than the predictions in the six scenarios with testing. In the scenarios without testing, the weekly number of infected increases near the end of the semester (as shown in Figure 6).

The predictions of the total number of infected varied by scenario and model, as shown in Table 2. The predictions from the ABM and DCM were more similar in the scenarios without testing (the odd-numbered scenarios) than in the scenarios with testing (the even-numbered scenarios).

Overall, the predictions by the ABM and DCM are more similar to each other, and they are less similar to the predictions from the ODEM.

This result also held for the modified models with the student cohorts (described in Section 2.6). The most notable changes were a much smaller fluctuation in the predicted number of persons in isolation and quarantine and a smaller value for the predicted number of those who become infected and then recover.

**Table 2.** Total number of persons infected by scenario and model. For the ABM, the value given is the average of ten replications; the value in parentheses is the standard deviation).

| Scenario | ABM | DCM | ODEM |
|---|---|---|---|
| 1 | 1131 (93) | 1055 | 1335 |
| 2 | 297 (17) | 46 | 95 |
| 3 | 703 (47) | 652 | 118 |
| 4 | 406 (28) | 169 | 71 |
| 5 | 524 (41) | 466 | 148 |
| 6 | 201 (19) | 31 | 76 |
| 7 | 807 (51) | 740 | 171 |
| 8 | 296 (16) | 45 | 78 |
| 9 | 361 (15) | 398 | 96 |
| 10 | 159 (15) | 26 | 52 |
| 11 | 211 (15) | 208 | 52 |
| 12 | 102 (7) | 17 | 41 |

*3.2. Parsimony*

Overall, the three models required the same set of input data: the characteristics of the population, the disease progression, the probability of transmission and the factors that affect it, and the intervention scenarios. The models used this common data in different ways and created different types of outputs, however.

The ODEM used the input data to determine its specific inputs: the values for eight parameters (coefficients), the population size, and the initial number of infected. For each scenario, it yielded a set of timestamps (the average over 12 scenarios was 181 timestamps) and nine compartment sizes for each time step (on average, 1635 values). (The timestamps that were determined when solving the system of differential equations did not correspond exactly to the days in the time horizon.)

The DCM yielded twenty compartment sizes for each time step (119 time steps; 2380 values).

The ABM tracked the state of each person using ten variables. For each replication, the model yielded the state of each person at the end of the time horizon (14,000 values) and the number of persons in six categories for each time step (119 time steps; 714 values).

### 3.3. Computational Effort

To measure the computational effort, we used the "Run and Time" function in MATLAB; the reported computational time did not include any time for reading input data from files, printing messages, or drawing graphs.

Running ten replications of the ABM required 14,762 seconds in the baseline scenario. Updating and accessing sparse matrices of contacts between persons required 14,068 seconds (95.3% of the time); the remaining steps required 694 seconds (4.7% of the time).

The DCM required 0.026 seconds in the baseline scenario. The ODEM required 0.040 seconds in the baseline scenario.

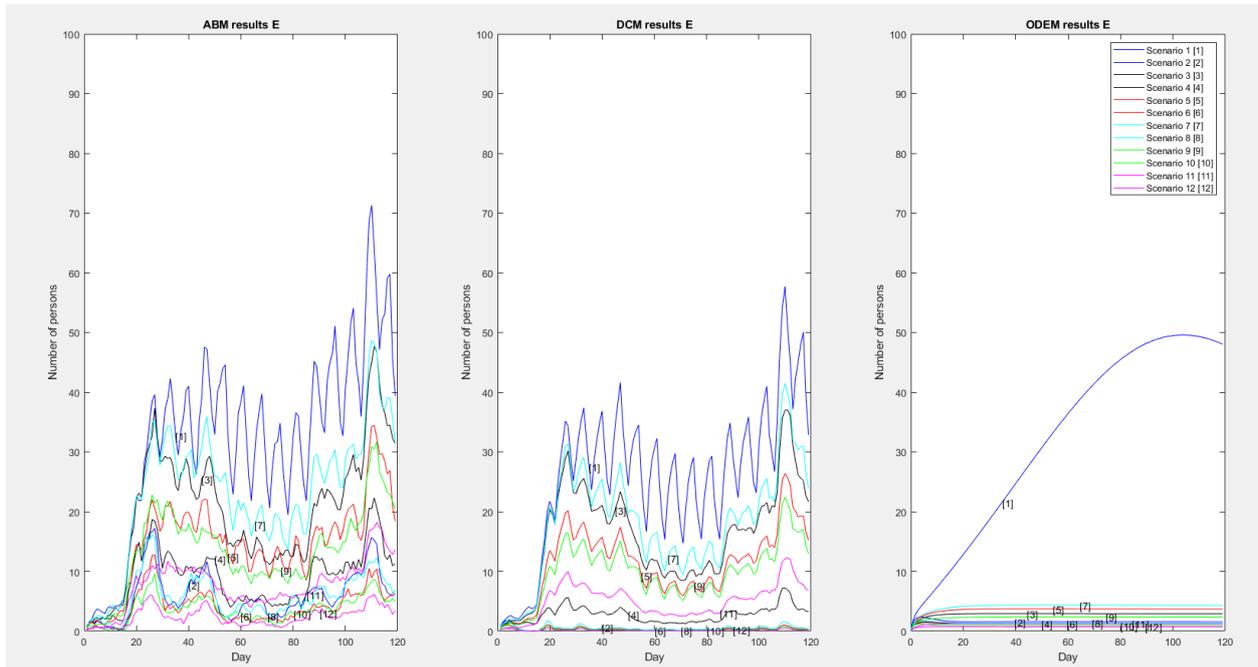

**Figure 3.** The predicted number of exposed (not in isolation or quarantine) at the end of each day of the time horizon from the ABM, DCM, and ODEM in scenarios 1 to 12.

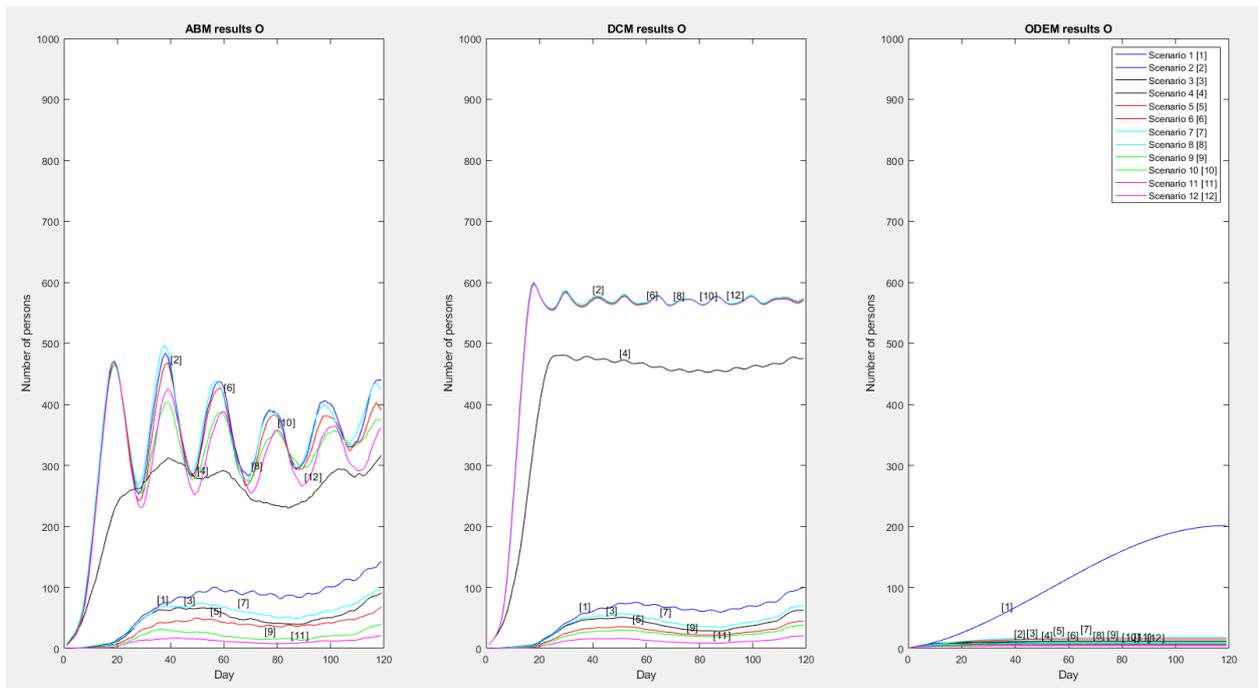

**Figure 4.** The predicted number of persons in isolation at the end of each day of the time horizon from the ABM, DCM, and ODEM in scenarios 1 to 12.

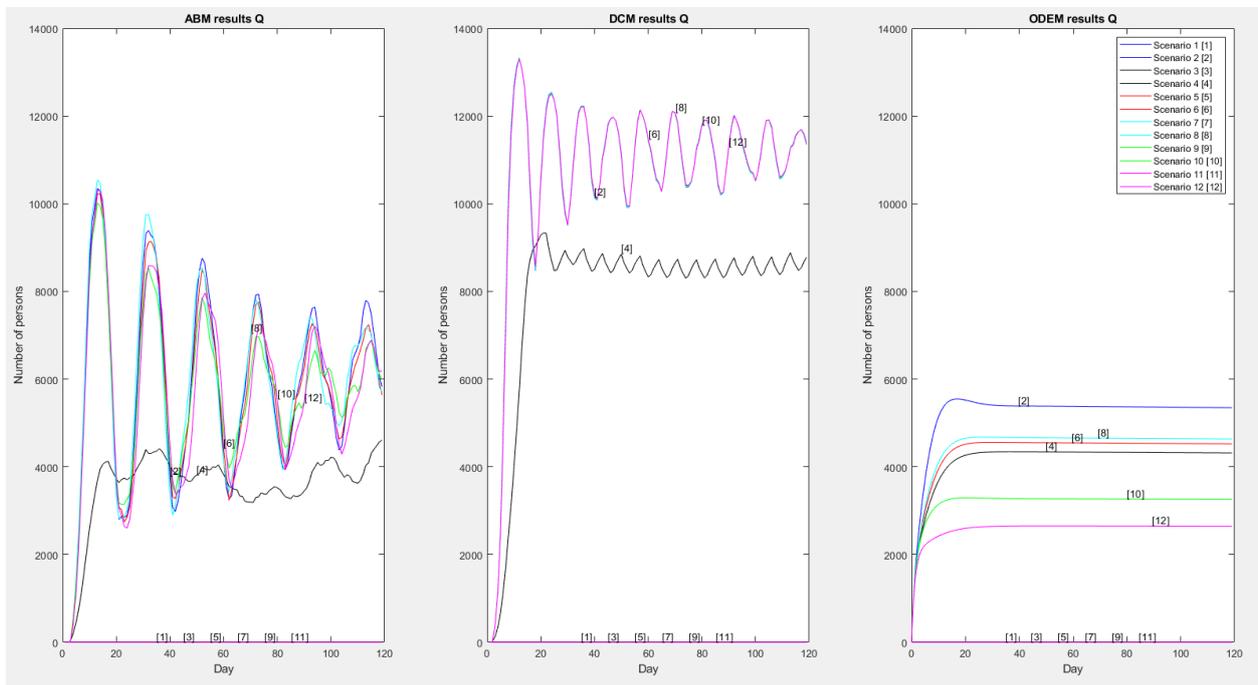

**Figure 5.** The predicted number of persons in quarantine at the end of each day of the time horizon from the ABM, DCM, and ODEM in scenarios 1 to 12.

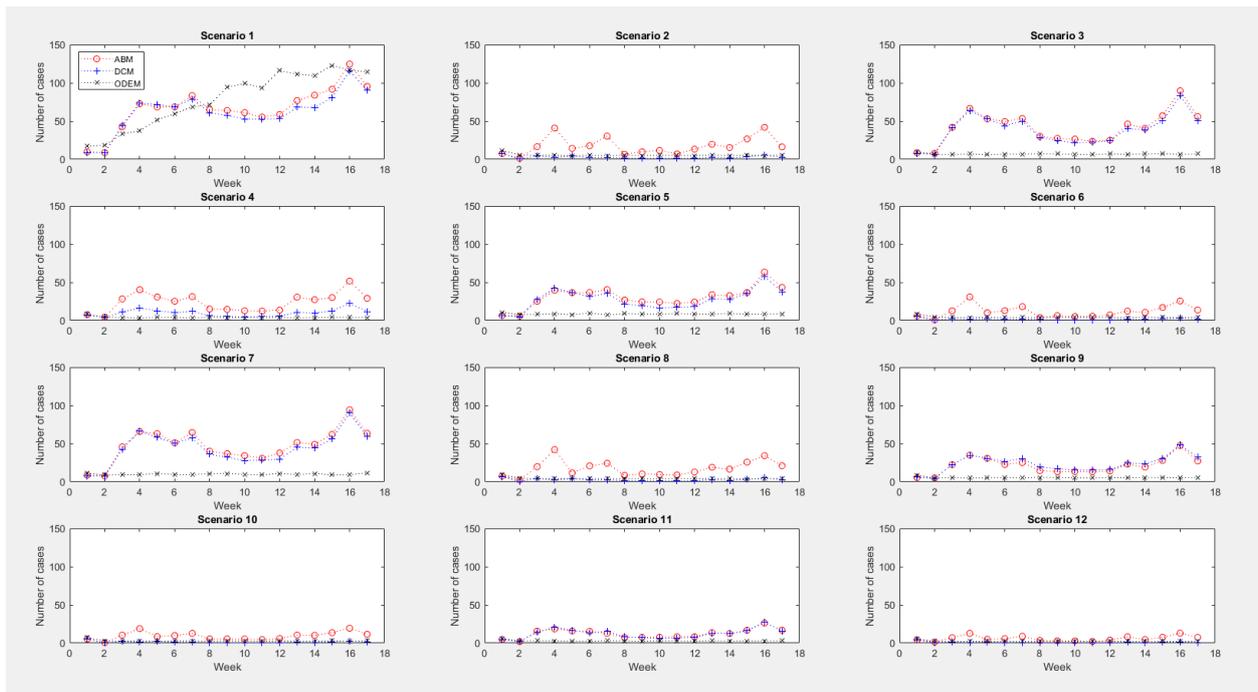

**Figure 6.** The predicted number of new cases during each week of the time horizon from the ABM, DCM, and ODEM in scenarios 1 to 12.

## 4. Discussion

A key feature of the ABM is the ability to model each person. In our study, this meant that, in the ABM, some specific persons wore masks and some specific persons were vaccinated, for example. It would be possible to use detailed data about each student's courses, residence, and activities and data about each instructor's course. (Such data was not available for this study, however.) Moreover, it would be possible to model characteristics such as age and pre-existing health conditions that might affect the progression of the disease for each person. Finally, the model gives the analyst the ability to track the spread of the disease from person to person within the community. This power requires more computational effort and resources, however, including memory to track the state of each person and the interactions between persons. Using sparse matrices and other techniques can reduce this somewhat, but the difference between the ABM's computation time and the other two models' computation time is vast.

In addition, the ABM is a stochastic model and can provide insights into the distribution of possible outcomes; the other two models focus on expected values. Considering the range of outcomes might be useful for planning stockpiles or adding resources as buffers for the peak demand. In some cases, because real-world data were not available, the ABM used random processes (such as class selection or vaccination) that don't match how these processes are done in practice.

At the other extreme, the ODEM aggregates many of the details in order to determine the coefficients for nine equations: one for each compartment. The model's coefficients are the same throughout the time horizon; it considers how the number of persons in each compartment changes over time, but it ignores how the coefficients might change. This model also combines students and faculty. These features of the model significantly affect its predictions, which are unlike those from the ABM and DCM. Although the magnitudes of the numbers are close, the periodic changes don't appear. Thus, it seems that the ODEM failed to capture some significant dynamics.

The DCM appears to be a desirable compromise. By considering only compartments (not individual persons), its computational burden is much less than that of the ABM. By including factors that change from day to day, it seems to capture some of the dynamics that are present in the ABM's predictions. Although its predictions to do not always match those from the ABM, they clearly show the importance of testing and reducing the number of courses that meet on campus and the much smaller impact of other interventions.

One might modify these models with a more detailed disease progression model, details about which students lived with whom, other activities on campus, data about students who lived off campus, and similar factors. It seems unlikely, however, that including these details would change the key differences between these models. The details of the equations in the DCM and values of the coefficients in the ODEM would change, but these models would still run very quickly and provide only aggregate results. The ABM would become even more complicated, and running it would still require more time and memory.

In this study, the models' predictions are useful only for comparing the scenarios to each other and comparing the models, which was our objective. This study did not try to model the actual conditions in the community, and we did not compare the models' predictions to the actual spread of the disease. Thus, we cannot make any claims about which model is most "accurate."

Instead, our goal was to compare the models, and these results allow us to do that. Any factors that do not match actual characteristics of the disease or the community are constant across the models and should not affect our comparison of the models. It seems unlikely that any model can perfectly model the real-world pandemic at any scale [23].

For our goal, we were satisfied to create a synthetic population and include realistic factors about the disease and the impact of the interventions without compelling our models to meet criteria that were not relevant to this study.

Although model validation is ultimately a matter of judgment [24], we have worked to justify the models' validity in multiple ways, as the literature suggests [25]. First, regarding the models' conceptual validity, one author created the models, and the other two authors reviewed the models and agreed that they were reasonable for the scenarios that our study considered (so this rests on the authors' combined expertise in infectious diseases, modeling, and epidemiology). The SEIR disease progression has been used in numerous studies of COVID-19, and all three modeling techniques have been used in previous studies.

Regarding data validity, we believe that the model's input values reasonably reflect actual real-world values. Where possible we used values based on data about testing at the University of Maryland, the length of the semester at the university, and parameters reported in the literature. In some cases, however, we needed to estimate values and make assumptions due to a lack of data, but, because we used the same values in all three models, we believe that our results are not sensitive to these particular values.

Finally, the models' predictions matched reasonably well the results of other studies that have modeled disease spread and our beliefs about the impact of different interventions.

## 5. Summary and Conclusions

This paper described a study to compare three models that can estimate the trajectory of a COVID-19 outbreak in a university community and predict the impact of different public health interventions. This study considered twelve scenarios that included a baseline scenario (with no interventions) and other scenarios with combinations of testing, moving courses off-campus, masks, improvements to classroom environment, and vaccination. The results showed that testing and moving courses off-campus had the most impact on the spread of the disease over one semester. This illustrates the utility of such models to decision-makers who need to evaluate the relative effectiveness of different interventions such as mask mandates.

These results yielded some specific insights into the tradeoffs involved when selecting a model to predict a disease outbreak. The agent-based model allowed us to consider factors at the level of individual persons, but this model required the most computational effort and memory. The details about each person may be interesting but don't provide more information about the impact of different interventions. The ODEM, which uses differential equations and is deterministic, aggregates many factors and uses less computational resources but fails to capture some dynamics that affect the disease spread. The DCM may be the best choice, as it requires less computational resources but still allows one to capture key dynamics.

More generally, this study revealed the difficulties of modeling disease spread in a particular community. Constructing a highly accurate model would require much data about the persons in the community, which may be difficult to obtain due to the need to protect their privacy. Using a synthetic population avoids this challenge, but the resulting predictions may be inaccurate, and it is a challenge to know how much error is introduced.

Still, the benefits of a deterministic compartment model should be relevant in many domains. This type of model can explicitly incorporate many details, especially those that change from day to day (such as class schedules) and quantities that describe what has happened previously. The compartments are not limited to the traditional compartments of a disease progression model (such as SEIR); they can represent combinations of states (such as isolation or quarantine and disease progression) and still model a large population with a much smaller number of compartments.

Ultimately, models like those that we studied can estimate the trajectory of a COVID-19 outbreak and predict the impact of different public health interventions and can help university administrators and public health officials evaluate their options and select the best strategies to respond to the challenges of COVID-19.

Protecting public health during the COVID-19 pandemic is a key national priority at the current time. Moreover, additional pandemics of contagious diseases may occur in the future. Model-based risk management will support the design and implementation of effective risk management strategies that mitigate disease spread and the related morbidity and mortality while allowing organizations to maintain some operations, which reduces the economic impact of the pandemic. The risk management and decision-making approaches that we develop for a university community can be adapted for other type of communities as well.

**Acknowledgments**

This research was supporting by the National Science Foundation grant CMMI2034755. This study used data collected by Oghenetekevwe Akoroda and Samuel Langlois.

**Authors' contributions**

J.W. Herrmann: study design, model creation, data analysis, paper writing and review. H. Liu: study design, paper writing and review. D.K. Milton: study design, paper writing and review.

**Conflict of interest**

The authors declare there is no conflict of interest.

**Appendix A**

This appendix describes the notation and equations for the deterministic compartment model (DCM). The equations presented here describe the general model that can be used for all of the scenarios that were considered in this study. The final part of this appendix describes the specific settings used for the different scenarios.

The DCM included twenty compartments, ten for students and ten for faculty:

- Susceptible (S)
- Exposed (E)
- Infectious (I)
- Recovered (R)
- Susceptible in isolation (SI)
- Exposed in isolation (EI)
- Infectious in isolation (II)
- Susceptible in quarantine (SQ)
- Exposed in quarantine (EQ)
- Infectious in quarantine (IQ)

In this notation, the subscript $a$ will be an element of the set $\boldsymbol{S}$ = {S, E, I, R, SI, EI, II, SQ, EQ, IQ}, and the superscript $v$ will designate the group: group 1 includes the students, and group 2 is the faculty.

Let $n_a^v(t)$ be the expected number of persons in group $v$ in compartment $a$ at the end of day $t$. Let $f_{ab}^v(t)$ be the expected number of persons in group $v$ who move from compartment $a$ to compartment $b$ during day $t$. $f_{aa}^v(t) = 0$ for all $a$ in $\boldsymbol{S}$, $v$ = 1, 2, and all $t$.

For ease of implementation, we defined intermediate quantities $y_a^v$ that are recalculated each day to determine the transitions that day. Because they are intermediate quantities, the notation here is simplified by ignoring the day.

Let $N^v$ be the total number of persons in each group, $v$ = 1, 2.

Let $N_0$ be the initial number of persons who are infectious.

Let $p_{vax}(t)$ be the proportion of vaccinated students and faculty on day $t$.

Let $p_{mi}$ be the proportion of students and faculty on day $t$ who are wearing a mask of type $i$.

Let $p_{is}$ be the proportion of infectious persons who show symptoms (and are tested).

Let $t_{test}$ be the testing period (days).

Let $t_{inc}$ be the incubation period (days).

Let $t_{rec}$ be the recovery period (days).

Let $t_{iso}$ be the isolation period (days).

Let $t_q$ be the quarantine period (days).

Let $t_d$ be the delay in contact tracing (days).

Let $t_b$ be the duration of the time horizon for contact tracing (days).

Let $n_{cs}$ be the number of classes that each student takes.

Let $s_{cs}$ be the number of students in each class.

Let $C(t)$ = 1 if classes are held on day $t$ and 0 otherwise.

Let $\Phi_{CS}$ be the fraction of classes that are meeting.

Let $\Phi_{CC}$ be the fraction of classes that are meeting in upgraded (cleaner) classrooms.

Let $n_{CT}$ be the number of contacts that a person has on-campus each day.

Let $N_{comm}(t)$ be the number of contacts per person on day $t$ with infectious members of the community off-campus.

Let $P_a^+$ be the probability of a positive COVID-19 test result for persons in compartment $a$ in **S**.

Let $P_{comm}^{TR}$ be the probability of transmission per contact with an infectious community member off-campus.

Let $P_{cp}^{TR}$ be the probability of transmission per contact with an infectious student or faculty.

Let $c_{vax}^{TR}$ be the coefficient that reduces the likelihood of becoming infected for someone who is vaccinated.

Let $c_{mi}^{TR}$ be the coefficient that reduces the probability of transmission due to a mask of type $i$ ($i = 1, 2$).

Initialization. For $v = 1, 2$:

$$n_I^v(0) = N_0 N^v / (N^1 + N^2)$$

$$n_S^v(0) = N^v - n_I^v(0)$$

Periodic testing ($v = 1, 2, a = $ S, E):

$$f_{a,aI}^v(t) = P_a^+ n_a^v(t-1) / t_{test}$$

$$y_a^v = n_a^v(t-1) - f_{a,aI}^v(t)$$

Symptomatic and periodic testing of infectious ($v = 1, 2$):

$$f_{I,II}^v(t) = P_I^+ p_{is} f_{E,I}^v(t-1) + P_I^+ \left( n_I^v(t-1) - f_{E,I}^v(t-1) \right) / t_{test}$$

$$y_I^v = n_I^v(t-1) - f_{I,II}^v(t)$$

Daily testing of persons in quarantine ($v = 1, 2, a = $ S, E, I):

$$f_{aQ,aI}^v(t) = P_a^+ n_{aQ}^v(t-1)$$

$$y_{aQ}^v = n_{aQ}^v(t-1) - f_{aQ,aI}^v(t)$$

Quarantine contacts of those who were isolated previously:

Let $N_I^v$ be the number of persons who were previously isolated:

$$N_I^v = F_{SI}^v(t - t_d) + F_{EI}^v(t - t_d) + F_{II}^v(t - t_d)$$

For $v = 1, 2$, let $p_{CT}^v(\tau)$ be the fraction of contacts on day $\tau$ who were in group $v$, and let $p_a^v(\tau)$ be the fraction of contacts in group $v$ who were in state $a$ on day $\tau$ ($a = $ S, E, I):

$$G^v(\tau) = n_S^v(\tau) + n_E^v(\tau) + n_I^v(\tau) + n_R^v(\tau)$$

$$p_{CT}^v(\tau) = G^v(\tau) / \left( G^1(\tau) + G^2(\tau) \right)$$

$$p_a^v(\tau) = n_a^v(\tau) / G^v(\tau)$$

For $v = 1, 2$, $w = 1, 2$, and $a$ in {S, E, I}, let $N_{Cwa}^v$ be the number of contacts (over the time horizon for contact tracing) in state $a$ in group $w$ of a person in group $v$ who was previously isolated:

$$N_{C1a}^1 = \sum_{\tau = t - t_d - t_b}^{t-1} p_a^1(\tau) p_{CT}^1(\tau) n_{CT} + C(t) p_a^1(\tau) \Phi_{CSc} n_{cs} (s_{cs} - 1) / 2$$

$$N_{C2a}^1 = \sum_{\tau = t - t_d - t_b}^{t-1} p_a^2(\tau) p_{CT}^2(\tau) n_{CT} + C(t) p_a^2(\tau) \Phi_{CS} n_{cs} / 2$$

$$N_{C1a}^2 = \sum_{\tau = t - t_d - t_b}^{t-1} p_a^1(\tau) p_{CT}^1(\tau) n_{CT} + C(t) p_a^1(\tau) \Phi_{CS} s_{cs} / 2$$

$$N_{C2a}^2 = \sum_{\tau = t - t_d - t_b}^{t-1} p_a^2(\tau) p_{CT}^2(\tau) n_{CT}$$

For $v = 1, 2$ and $a$ in {S, E, I}, the number of people who begin quarantine depends upon the number who were previously isolated and the number of contacts that they had:

$$f_{a,aQ}^v(t) = N_I^1 N_{Cva}^1 + N_I^2 N_{Cva}^2$$

Update number in quarantine and isolation ($v = 1, 2$; $a =$ S, E, I):

$$y_{aQ}^v = n_{aQ}^v(t-1) + f_{a,aQ}^v(t) - f_{aQ,aI}^v(t)$$

$$F_{aI}^v(t) = f_{a,aI}^v(t) + f_{aQ,aI}^v(t)$$

$$y_{aI}^v = n_{aI}^v(t-1) + F_{aI}^v(t)$$

After removing those who are isolated and quarantined, let $p_{INF}^v$ be the fraction of contacts in group $v$ who are infectious, and let $p_A^v$ be the fraction of contacts who are in group $v$.

$$p_{INF}^v = y_I^v / \left( y_S^v + y_E^v + y_I^v + y_R^v \right)$$

$$p_A^v = \left( y_S^v + y_E^v + y_I^v + y_R^v \right) / \sum_{w=1}^{2} \left( y_S^w + y_E^w + y_I^w + y_R^w \right)$$

For $v = 1, 2$ and $w = 1, 2$, let $N_{SICw}^v$ be the number of infectious contacts from group $w$ of a person in group $v$ who is susceptible this day:

$$N_{SIC1}^1 = p_{INF}^1 p_A^1 n_{CT} + C(t) p_{INF}^1 \Phi_{CS} n_{cs} (s_{cs} - 1) / 2$$

$$N_{SIC2}^1 = p_{INF}^2 p_A^2 n_{CT} + C(t) p_{INF}^2 \Phi_{CS} n_{cs} / 2$$

$$N_{SIC1}^2 = p_{INF}^1 p_A^1 n_{CT} + C(t) p_{INF}^1 \Phi_{CS} s_{cs} / 2$$

$$N_{SIC2}^2 = p_{INF}^2 p_A^2 n_{CT}$$

Let $c_m$ be the coefficient that reduces the probability of transmission based on the proportion of students and faculty who are wearing masks:

$$c_m = (1 - p_{m1} - p_{m2} + p_{m1}c_{m1}^{TR} + p_{m2}c_{m2}^{TR})$$

The number of susceptible who become infected (and move to exposed) ($v = 1, 2$):

$$f_{S,E}^v(t) = y_S^v(1 - p_{vax}(t) + p_{vax}(t)c_{vax}^{TR})\left(c_m^2 P_{cp}^{TR}(N_{SIC1}^v + N_{SIC2}^v) + c_m P_{comm}^{TR} N_{comm}(t)\right)$$

The number of exposed who become infectious ($v = 1, 2$):

$$f_{E,I}^v(t) = y_E^v / t_{inc}$$

The number of infectious who recover ($v = 1, 2$):

$$f_{I,R}^v(t) = y_I^v / t_{rec}$$

The number of susceptible in isolation who leave isolation ($v = 1, 2$):

$$f_{SI,S}^v(t) = f_{S,SI}^v(t - t_{iso})$$

The number of exposed in isolation who become infectious ($v = 1, 2$):

$$f_{EI,II}^v(t) = y_{EI}^v / t_{inc}$$

The number of infectious in isolation who recover ($v = 1, 2$):

$$f_{II,R}^v(t) = y_{II}^v / t_{rec}$$

The number of susceptible in quarantine who leave isolation is estimated by determining the fraction of those currently in quarantine who arrived $t_q$ days ago ($v = 1, 2$):

$$f_{SQ,S}^v(t) = y_{SQ}^v f_{S,SQ}^v(t - t_q) / \sum_{\tau = t - t_q}^{t-1} f_{S,SQ}^v(\tau)$$

The number of exposed in quarantine who become infectious ($v = 1, 2$):

$$f_{EQ,IQ}^v(t) = y_{EQ}^v / t_{inc}$$

The number of infectious in quarantine who leave recover ($v = 1, 2$):

$$f_{IQ,R}^v(t) = y_{IQ}^v / t_{rec}$$

Update the number of persons in each compartment ($v = 1, 2$, $a$ in $\boldsymbol{S}$):

$$n_a^v(t) = n_a^v(t-1) + \sum_{b \in S} f_{ba}^v(t) - \sum_{b \in S} f_{ab}^v(t)$$

The following values are set for all scenarios: $N^1$ = 12,000 students. $N^2$ = 2,000 faculty. $N_0$ = 3 persons. $p_{is}$ = 0.6, and $t_{iso}$ = 10 days of isolation. $n_{cs}$ = 5 courses, and $s_{cs}$ = 30 students in each course. $C(t)$ = 1 every Monday, Tuesday, Wednesday, and Thursday and 0 otherwise. $n_{CT}$ = 4 contacts per day on campus. $P_a^+$ = 0.99 for persons who are infectious; $P_a^+$ = 0.95 for persons who are exposed; and $P_a^+$ = 0.005 for persons who are susceptible. $P_{comm}^{TR}$ = 0.00004. $P_{cp}^{TR}$ = 0.0025.

**Periodic testing.** In scenarios without periodic testing and quarantine, $f_{a,aI}^{v}(t) = 0$ for $v = 1, 2$, and $a = $ S,

E. $f_{I,II}^{v}(t) = P_{I}^{+} p_{is} f_{E,I}^{v}(t-1)$ for $v = 1, 2$. For $v = 1, 2$ and $a$ in {S, E, I}, $f_{a,aQ}^{v}(t) = 0$.

In scenarios with periodic testing and quarantine, $t_{test} = 14$ days, and $t_q = 10$ days. For contact tracing, $t_d = 1$ day, and $t_b = 5$ days.

**On-campus courses.** In scenarios with all on-campus courses, $\Phi_{CS} = 1$. In scenarios with fewer on-campus courses, $\Phi_{CS} = 0.20$.

**Masks.** We included two types of masks: "tight" (type 1) and "loose" (type 2). In scenarios without masks, $p_{m1} = p_{m2} = 0$. In scenarios with masks, $p_{m1} = p_{m2} = 0.25$. $c_{m1}^{TR} = 0.1$, and $c_{m2}^{TR} = 0.5$.

**Scenarios with environmental improvements to classrooms.** In scenarios without improvements, $\Phi_{CC} = $

0. In scenarios with improvements, $\Phi_{CC} = 1$.

**Vaccination.** In scenarios without vaccination, $p_{vax}(t) = 0$ for all $t$. In scenarios with vaccination, $p_{vax}(t) = 0.8$ for all $t$. $c_{vax}^{TR} = 0.4$.

## Appendix B

This appendix describes the equations for the ODEM and how the coefficients are calculated from the input values.

Let $S(t)$, $E(t)$, $I(t)$, and $R(t)$ be the number susceptible, exposed, infectious, and recovered at time $t$. Let $EI(t)$ and $II(t)$ be the number of exposed and infectious in isolation at time $t$. Let $SQ(t)$, $EQ(t)$, and $IQ(t)$ be the number of susceptible, exposed, and infectious in quarantine at time $t$.

Let $\beta_1$ and $\beta_2$ be the parameters that determine the spread of the disease. Let $\gamma$ and $\lambda$ be the coefficients that describe the progress of the disease; let $\tau$ be the rate of testing those who have not recovered. The differential equations for this model were the following:

$$\frac{dS}{dt} = -\beta_1 S(t)I(t) - \beta_2 S(t) - \alpha_1(\tau_1 E(t) + \tau_2 I(t))S(t) / (S(t) + E(t) + I(t) + R(t)) + \alpha_2 SQ(t)$$

$$\frac{dE}{dt} = \beta_1 S(t)I(t) + \beta_2 S(t) - (\gamma + \tau_1)E(t) - \alpha_1(\tau_1 E(t) + \tau_2 I(t))E(t) / (S(t) + E(t) + I(t) + R(t))$$

$$\frac{dI}{dt} = \gamma E(t) - (\tau_2 + \lambda)I(t) - \alpha_1(\tau_1 E(t) + \tau_2 I(t))I(t) / (S(t) + E(t) + I(t) + R(t))$$

$$\frac{dEI}{dt} = \tau_1 E(t) + (1 - \gamma)EQ(t) - \gamma EI(t)$$

$$\frac{dII}{dt} = \gamma EI(t) + \tau_2 I(t) + (1 - \lambda)IQ(t) - \lambda II(t)$$

$$\frac{dR}{dt} = \lambda I(t) + \lambda II(t) + \lambda IQ(t)$$

$$\frac{dSQ}{dt} = \alpha_1(\tau_1 E(t) + \tau_2 I(t))S(t) / (S(t) + E(t) + I(t) + R(t)) - \alpha_2 SQ(t)$$

$$\frac{dEQ}{dt} = \alpha_1(\tau_1 E(t) + \tau_2 I(t))E(t) / (S(t) + E(t) + I(t) + R(t)) - \gamma EQ(t) - (1 - \gamma)EQ(t)$$

$$\frac{dIQ}{dt} = \alpha_1(\tau_1 E(t) + \tau_2 I(t))I(t) / (S(t) + E(t) + I(t) + R(t)) + \gamma EQ(t) - \lambda IQ(t) - (1 - \lambda)IQ(t)$$

Based on previous of studies of COVID-19 outbreaks (Benneyan *et al.*, 2021), we set $\lambda = 1/14$, $\gamma = 1/3$, and, after calibration, set the baseline $\beta_1 = \beta_2 = 0.00005$. We set $P_{comm}^{TR} = 0.00004$, and $P_{cp}^{TR} = 0.0025$.

If surveillance testing is not active, then $\tau_1 = 0$ and $\alpha_1 = 0$. Else, $\tau_1 = 1/14$ and $\alpha_1 = 1000$. $\tau_2 = 0.6$. $\alpha_2 = 1/10$.

Based on the interventions for each scenario, we modified the value of the $\beta_1$ parameter in this model by multiplying it by the following factors.

Let $\Phi_{CS}$ be the fraction of courses that are meeting in-person on campus. In most scenarios, $\Phi_{CS} = 1$. In scenarios 3 and 4, however, $\Phi_{CS} = 0.2$. Let $c_A$ be the coefficient related to in-person courses.

$$c_A = \frac{N^1\left(n_{CT} + \frac{4}{7}\Phi_{CS}n_{cs} / 2\right)(s_{cs} + 1) + N^2\left(n_{CT} + \frac{2}{7}\Phi_{CS}s_{cs}\right)}{N^1\left(n_{CT} + \frac{4}{7}n_{cs} / 2\right)(s_{cs} + 1) + N^2\left(n_{CT} + \frac{2}{7}s_{cs}\right)}$$

We adjusted this model for scenarios with mask-wearing as follows: Let $N_{pop}$ be the total number of persons in the population. Let $N_{TM}$ be the total number who wear tight-fitting masks. Let $N_{LM}$ be the total number who wear loose-fitting masks. Then, define the following fractions that describe the fraction of the population who wear no masks, who wear tight-fitting masks, and who wear loose-fitting masks:

$$f_0 = (N_{pop} - N_{TM} - N_{LM}) / N_{pop}$$

$$f_1 = N_{TM} / N_{pop}$$

$$f_2 = N_{LM} / N_{pop}$$

Let $c_B$ be the coefficient related to masks:

$$c_B = (f_0 + 0.1f_1 + 0.5f_2)^2$$

Let $c_C$ be the coefficient related to cleaner classrooms:

$$r = \left(1 - \sqrt{1 - P_{cp}^{TR}}\right) / P_{cp}^{TR}$$

$$c_C = 1 + (r - 1)\Phi_{CC}$$

Let $c_D$ be the coefficient related to vaccination:

$$c_D = 1 + \left(c_{vax}^{TR} - 1\right) p_{vax}$$

After calculating these, we modified the parameter values in the model as follows to reduce the force of infection:

$$\beta_1^* = \beta_1 c_A c_B c_C c_D$$

$$\beta_2^* = \beta_2 c_D$$